\documentclass[%
 reprint,
superscriptaddress,
 amsmath,amssymb,
 aps,
]{revtex4-2}
\usepackage{tikz}
\usepackage{graphicx}
\usepackage{dcolumn}
\usepackage{bm}

\usepackage{wasysym}
\usepackage{oplotsymbl}

\graphicspath{{./Fig_Pe_Dr/}{./}} 

\begin{document}

\preprint{APS/123-QED}

\title{Binary Mixtures of Intelligent Active Brownian Particles with Visual Perception}
\thanks{Electronic supplement available }%

\author{Rajendra Singh Negi}
\email{ranegi@syr.edu}
\affiliation{Theoretical Physics of Living Matter, Institute for Advanced Simulation, Forschungszentrum J{\"u}lich, 52425 J{\"u}lich, Germany}
\affiliation{Department of Physics, Syracuse University, Syracuse, New York 13244, USA}

\author{Roland G. Winkler}
\email{rg\_winkler@gmx.de}
\affiliation{Theoretical Physics of Living Matter, Institute for Advanced Simulation, Forschungszentrum J{\"u}lich, 52425 J{\"u}lich, Germany}

\author{Gerhard Gompper}
\email{g.gompper@fz-juelich.de}

\affiliation{Theoretical Physics of Living Matter, Institute for Advanced Simulation, Forschungszentrum J{\"u}lich, 52425 J{\"u}lich, Germany}

\date{\today}

\begin{abstract}

The collective properties of a binary mixture of $A$- and $B$-type self-steering particles endowed with visual perception are 
studied by computer simulations. Active Brownian particles are employed with an additional steering mechanism, which enables 
them to adjust their 
propulsion direction relative to the instantaneous positions of neighboring particles, depending on the species, either steering 
toward or away from them. Steering can be nonreciprocal between the $A$- and $B$-type particles.
The underlying dynamical and structural properties of the system are governed by the strength and polarity of the maneuverabilities associated with the vision-induced steering. The model predicts the emergence of a large variety of 
nonequilibrium behaviors, which we systematically characterize for all nine principal sign combinations of $AA$, 
$BB$, $AB$ and $BA$ maneuverabilites. In particular, we observe the formation of multimers, encapsulated aggregates, 
honeycomb lattices, and predator-prey pursuit. 
Notably, for a predator-prey system, the maneuverability and vision angle employed by a predator significantly 
impacts the spatial distribution of the surrounding prey particles. For systems with electric-charge-like interactions  
and non-stochiometric composition, we obtain at intermediate activity levels an enhanced diffusion compared to 
non-steering active Brownian particles.  

\end{abstract}

\maketitle


\section{\label{sec:level1} Introduction}

Active-matter systems, ranging from microscopic biological systems such as bacteria  \cite{berg_1975_Nature, be_2020_CP, aranson_2022_RoPP} 
to macroscopic biological systems as flocks of birds \cite{cavagna_2014_ARCMP, pearce_2014_PNAS, vicsek_1995_PRL}, are intrinsically out of equilibrium and show a vast variety of fascinating emergent 
behaviors \cite{elgeti_2015_RPP, bechinger_2016_RMP, gompper_JPCM_2020, gompper_2025_roadmap}. 
Such features are not restricted to biological systems, but have been partially reproduced in artificial and hybrid 
systems, specifically in those comprised of Janus colloids \cite{park_2017_ACS, Tobias_2020_Nature_Qurom_sensing, bastos_2018_FRAI}. 
This makes active matter an important research field with significant biomedical potential, e.g., targeted drug delivery using 
engineered bacteria for tumor therapy \cite{yan_2020_AHM} and  biomedical untethered mobile milli/microrobots capable of accessing previously unreachable single cell sites for in situ and in vivo applications \cite{sitti_2015_IEEE}. It also offers promising solutions for environmental improvements, such as water treatment \cite{fu2_micromachine_2022,hu_2012_CSRev}.
Arising patterns and structures not only depend on the physical interactions between the various agents of an ensemble, but are often governed by nonreciprocal information input, e.g., directional visual perception in case of animals, processing of this information, and active self-steering response. 

Multi-component mixtures of self-steering active agents with nonreciprocal interactions can show an even more intriguing behavior. It is important to note that the nonreciprocal nature of interactions in multi-component
systems is on a different level compared to single-component systems, because the reaction of particles of one species 
to a particle of another species can be different from the inverse reaction 
\cite{you_2020_PNAS, Saha_2020_PRX, osat_2023_NN, meredith_2020_NC}.

Binary mixtures of active agents are prevalent in biological systems. For example, mixed swarms 
of bacteria \textit{Bacillus subtilis} and \textit{Pseudomonas aeruginosa} exhibit both cooperation and 
segregation across scales \cite{natan_2022_SR}. Mixtures of two strains of \textit{B.~subtilis} with 
distinct cell aspect-ratios show aggregate formation, in which longer cells serve as nucleation sites, attracting 
shorter and highly mobile cells \cite{wang_2004_molecular_M}. Mixing motile and nonmotile \textit{Escherichia coli} 
bacteria results in active density changes, where the circular movement of motile cells near surfaces 
creates flowing patterns that carry nonmotile cells along, while sedimentation disrupts vertical symmetry,
crucial for their gathering \cite{burriel2023active}.
Interactions between marine Pseudoalteromonas sp. and Gram-positive bacteria show predator-prey behavior, where 
the predator uses enzymes to kill Gram-positive cells while coexisting peacefully with Gram-negative bacteria, 
revealing how microbes can both compete and cooperate within mixed communities \cite{tang_NatComm_2020}. 
Recent experimental evidence indeed suggests that swimming and chemotactic sensing play an 
important role in shaping interbacterial interactions \cite{Seymour_2024_TrendsMicrobiol}. 
Furthermore, the relevance of sensing and self-steering is evident in macroscale binary biological systems. 
For instance, the flight of bee swarms to new nests consists of a binary mixture of 
streaker (leader) bees and follower (uninformed) bees, where bees use directed vision for attraction 
and repulsion among themselves \cite{fetecau_2012mathematical, beekman2006does}. Also, there are many 
examples of large animals hunting others, like wolves chasing deer or sharks hunting fish swarms.
In synthetic systems, chemotactic signaling in a mixture of microscale oil droplets of different 
chemistry embedded in micellar surfactant solutions have been shown to create predator-prey-like 
nonreciprocal chasing interactions \cite{meredith_2020_NC, sengupta_2011_PRE, cira_2015_Nature}.

Theoretical and simulation studies of model systems provide essential insight into the complex 
emergent behavior of active systems in general and binary mixtures of self-steering particles in particular.
They facilitate the characterization of the
emergent structures and dynamics and their dependence on the agent properties. Ultimately, they can provide a guide 
for the rational design of synthetic active agents with desired properties.  

``Dumb" active Brownian particles (ABPs) -- with conservative interactions only, like hard-core repulsion -- display aggregation in the form of motility-induced phase separation (MIPS) 
\cite{bialke_2012_PRL, wysocki_2014_EPL, cates_2015_ARCMP, pasquale_2018_PRL}  and nonequilibrium clustering in presence of hydrodynamic interactions \cite{peruani_2006_PRE, mario_2018_soft_matter}. Already binary mixtures of such active and passive Brownian particles show phase separation with domains enriched by passive or active particles, and propagating interfaces between them 
\cite{stenhammar_2015_PRL, wysocki_2016_NJP, wittkowski_2017_NJP, kolb_2020_SM}.
Experimental studies demonstrate that passive silica colloids rapidly self-assemble into tunable 2D structures via diffusiophoretic interactions with UV-activated Janus particles, with the clustering dynamics controlled by light intensity and particle size ratios \cite{singh_2017_AM}. Along the same line, simulations by varying the active particle density reveal a non-monotonic clustering trend, where high activity leads to an effective screening of phoretic interactions, slowing down aggregation \cite{jhajhria_2023_Soft_Mat}.

In systems with alignment of the propulsion direction with neighboring 
particles, as in the Vicsek model~\cite{vicsek_1995_PRL}, collective motion in single-component systems has 
been predicted in form of travelling bands \cite{vicsek_1995_PRL}. The incorporation of anisotropic sensory 
perception in that model significantly impacts the system's collective behavior \cite{gao_2025_PRR}. 
In single-component self-steering systems with pursuit-type steering, an even more complex behavior can 
emerge, such as milling \cite{negi_2024_PRR,saavedra_2024_PRL}, and the formation of
worm-like structures and worm-aggregate coexistence \cite{couzin_2002_Elsevier, barberis_2016_PRL, negi_2022_soft_matter, negi_2024_PRR, liu_2025_SM}. 
Variants of the  Vicsek model with two distinct species, for example systems of slow- and fast-moving self-propelled particles, where particles align within their own species and antialign with the other, yield micro-phase 
separation and parallel/antiparallel bands \cite{chatterjee_2023_PRE}, 
while mixtures of active and passive particles  with local alignment in confinement display segregation into an active 
core and a passive shell~\cite{agrawal_2021_PRE}. 

The emergent behavior becomes even more intriguing for active mixtures with nonreciprocal interactions between particle types. A prominent example is a mixture of chemically interacting particles, which produce or consume 
a chemical to which they are attracted or repelled corresponding to positive or negative chemotaxis.  
Such systems display a wealth of active clustering and phase-separation phenomena, such as association of particles into small molecules, aggregation into a static dense phase that coexists with a dilute phase, and the formation of non-stochiometric 
self-propelled macroscopic clusters with a comet-like tail \cite{Canalejo_2019_PRL}. 
Self-propulsion heterogeneity and nonreciprocity of binary interactions can drive the partial segregation of 
mixtures of active colloids \cite{maity_2023_PRL}. \\[1ex]

Here, we explore the collective behavior and structure formation of binary mixtures of active particles 
with visual perception and nonreciprocal self-steering. An agent is modeled as an ``intelligent'' active Brownian particle (iABP), with 
implicit self-steering abilities. An iABP can move toward or away from its own or the other species, with limited
maneuverability $\Omega$, by sensing the positions of neighboring particles in its vision cone, as illustrated in 
Fig.~\ref{fig:schematics}. Despite the considered minimal model, our approach contains 
a significant number of parameters, such as the activities of the two particle species, their vision angles and 
steering ability, the particle 
densities etc. This results in a rich phase behavior and interesting dynamics, which we are unable to  characterize 
completely here. Instead, we focus on various parameter combinations, which illustrate intriguing emergent behaviors.



%
%

\section{Modelling Binary Mixtures of Intelligent ABPs}
\label{sec:model_short}

\subsection{Equations of motion}

\begin{figure}
    \centering
    \includegraphics[width=.45\textwidth]{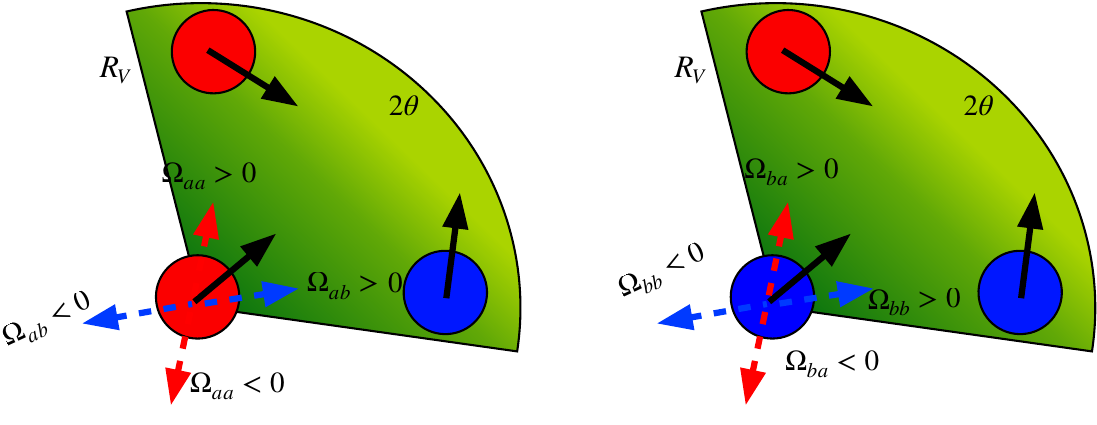}
    \caption{Schematic representation of the vision cone of a particle of the two types, A (red) and B (blue), and their corresponding 
    maneuverabilities. The dashed arrows indicate the nature of steering toward and away from detected particles within the vision cone.}
    \label{fig:schematics}
\end{figure}


In our minimalistic approach, the two types $A$ and $B$ of agents are described as self-steering active Brownian particles with visual perception. A mixture is composed of the total number $N= N_{A} + N_{B}$ of particles. The particle dynamics is governed by the  equations of motion \cite{das_2018_NJP, negi_2022_soft_matter,negi_2024_PRR}
\begin{equation} \label{eq:gen_eom_pos}
m \ddot{{\bm r}_i} = -\gamma_T \dot{{\bm r}_i}  + \gamma_T v_0 \bm{e}_i + \bm F_i +  \bm{\varGamma}_i(t), 
\end{equation}
where $\bm F^{a}_i(t) = \gamma_T v_0 \bm e_i(t)$ is the
propulsion force resulting in a speed $v_0$ along the direction $\bm e_i(t)$, which can change in response to the sensed environment.
Moreover, $m$ is the mass of an iABP, $\gamma_T$ the translational friction coefficient, and $\bm F_i$ the excluded-volume interactions between the iABPs. The latter are taken into  account by the truncated and shifted Lennard-Jones potential
\begin{equation} \label{eq:WCA_LGP}
    U_{LJ} = 
    \begin{cases}  \displaystyle 4 \epsilon \left[ {\left(\frac{ \displaystyle \sigma}{  \displaystyle r}\right)}^{12} - \left(\frac{ \displaystyle \sigma}{ \displaystyle r}\right)^{6} +1 \right] +\epsilon,       &  r \leq2^{1/6} \sigma 
    \\ 0 , &   r > \sigma  .
    \end{cases} 
\end{equation}
Here, $\sigma$ is the particle diameter, $\epsilon$ the repulsion strength, and $r$ the distance between two iABPs. 
Thermal fluctuations are considered as Gaussian and Markovian stochastic process $\bm \varGamma_i$  of zero 
mean and the second moments 
$\langle \bm\varGamma_i(t) \cdot \bm\varGamma_j(t') \rangle = 2d  \gamma_T k_BT \delta(t-t') \delta_{ij}$ in $d$ 
dimensions, with $T$ the temperature and $k_B$ the Boltzmann constant. 

An iABP is capability to respond to information regarding the position of the neighboring particles within a vision cone
by adjusting its propulsion direction. 
Figure~\ref{fig:schematics} illustrates the vision cones for a $A$- (red) particle and a $B$- (blue) particle, 
respectively, along 
with their respective capability (maneuverability) to respond to the instantaneous positions of their neighbors.

The evolution of the propulsion direction ${\bm{e}}^{\alpha}_i(t)$ 
of particle $i$ of type $\alpha$ ($\alpha, \beta$ $\in \{A,B\}$, $\alpha \neq \beta$) is governed by \cite{goh_2022_NJP, negi_2022_soft_matter}
\begin{equation} \label{eq:gen_force}
    \dot{\bm{e}}^{\alpha}_i(t)  =  \sum_{\gamma\in \{\alpha, \beta\}}\bm{M}^{\alpha \gamma}_i + \bm\varLambda_{i}(t) \times \bm{e}^{\alpha}_i(t) .
\end{equation}
The last term describes the (thermal) rotational diffusion independent of the particle type. It is modeled as Gaussian and Markovian stochastic processes of zero mean, variance  
$\langle \bm\varLambda_{i}(t) \cdot \bm\varLambda_{j} (t') \rangle = 2 (d-1)D_R \delta_{ij} \delta(t-t')$,
and the rotational diffusion coefficient $D_R$.

The cognitive ``visual'' torque $\bm M_i^{\alpha \gamma}$ by particles of the same  ($\gamma=\alpha$) or different ($\gamma=\beta$) type in the vision cone $VC$ is \cite{barberis_2016_PRL, negi_2022_soft_matter}
\begin{equation} \label{eq:f_aa}
     \bm{M}^{\alpha \gamma}_{i} = \frac{\Omega_{\alpha \gamma}}{N_{\alpha \gamma}} 
          \sum_{j\in VC}e^{-r_{ij}^{\alpha \gamma}/R_0} \bm{e}^{\alpha}_i \times ( \bm{u}_{ij}^{\alpha \gamma}\times \bm{e}^{\alpha}_i) .
\end{equation}
Here, $\bm u^{\alpha \gamma}_{ij} = \bm r_{ij}^{\alpha \gamma}/|\bm r_{ij}^{\alpha \gamma}|$ is the unit vector of the distance 
$\bm r_i^{\alpha} - \bm r_j^{\gamma}$ between particle $i$ and $j$ of the types $\alpha$ and $\gamma$, and 
\begin{align} \label{eq:eq_2_norm1}
  N_{\alpha \gamma} =  \sum_{j\in VC}e^{-r_{ij}^{\alpha \gamma}/R_0}
\end{align}
is the normalization factor, which is determined by the effective number of such particles in the vision cone.
The condition for particles $j$ to lie within the vision cone of particle $i$ is
\begin{equation}
	\bm{u}^{\alpha \gamma}_{ij} \cdot \bm e^{\alpha}_i \geq \cos \theta_\alpha ,
\end{equation}
where $\theta_\alpha$ -- the vision angle -- is the opening angle of the vision cone centered by the particle orientation 
$\bm e_{\alpha i}$, and $R_0$ describes the characteristic range of the visual perception. Additionally, the vision range is 
limited to 
\begin{equation}
   |\bm r_{i}^\alpha - \bm r_{j}^\gamma | \leq R_{v}, 
\end{equation}
treating all particles further apart than $R_v$ as invisible ($R_{v} \geq R_{0}$). In a dilute system, only a single 
particle may be within the distance $R_{V}$, and the exponential factor $ e^{-r_{ij}^{\alpha \gamma}/R_0}$ cancels out, 
as it appears both in the numerator and denominator in Eq. \eqref{eq:f_aa}. Conversely, in a dense system, an effective 
reduced vision range $R_0$ appears due to the exponential factor. This can be interpreted as blocking the view by 
neighboring particles highlighting the influence of the local environment on visual perception (similar to the
effect studied in Ref.~\cite{pearce_2014_PNAS}). 

The torque $\bm{M}^{\alpha \gamma}_{i}$ describes the preference of an iABP to move toward \cite{negi_2022_soft_matter, negi_2024_PRR}
or away \cite{negi_2024_SciRep, Iyer_2024_CommunPhys} from the center of mass of iABPs of type $\gamma$ in its vision cone $VC$, 
depending on the sign of the maneuverability $\Omega_{\alpha \gamma}$.
The normalization Eq.~\eqref{eq:eq_2_norm1} implies a nonadditive interaction, and emphasizes the importance of the effective 
particle number 
in the vision cone. It is motivated by the fact that additive and nonadditive interactions imply distinct macroscopic 
behaviors in interacting active systems \cite{chepizhko_2021_soft_matter}.

We focus here on two-dimensional systems. Then, with the representation of the propulsion direction in polar coordinates,
$\bm e_{i}^{\alpha} = (\cos \varphi_{i}^\alpha , \sin \varphi_{i}^\alpha)^T$ and the difference 
vector $\bm u_{ij}^{\alpha \beta} = (\cos \phi_{ij}^{\alpha \beta}, \sin \phi_{ij}^{\alpha \beta})^T$, Eq.~\eqref{eq:gen_force} 
implies the equation of motion for the orientation 
angle $\varphi_i^\alpha$
\begin{equation} \label{eq:GR_PC_A}
\begin{aligned}
     \dot{\varphi}_{i}^\alpha  & = 
      \sum_{\gamma \in \{\alpha, \beta\}} \frac{\Omega_{\alpha \gamma}}{N_{\alpha \gamma}} 
          \sum_{j\in VC}e^{-r_{ij}^{\alpha \gamma}/R_0} \sin({\phi_{ij}^{\alpha \gamma}-\varphi_{i}^\alpha}) \\
      & \hspace{4cm} + \Lambda_i(t) ,
  \end{aligned}
\end{equation}
with $\Lambda_i$ a Gaussian and Markovian stochastic process of zero mean and 
$\langle \Lambda_i(t) \Lambda_j(t') \rangle = 2  D_R \delta_{ij} \delta (t-t')$.

Although we consider a minimal model, the system contains a significant number of parameters, as 
there is the P{\'e}clet number $Pe=v_0/(\sigma D_R)$ (where $\sigma$ is the effective particle diameter) 
the vision angles $\theta_{\alpha}$ ($\alpha \in \{A,B\})$, the vision range 
$R_{0}$ and cutoff radius $R_v$, the  maneuverabilities $\Omega_{\alpha\beta}$,  the packing fraction 
$\Phi = \pi \sigma^2 N/(4L^2)$ (with linear systems size $L$), and the particle numbers $N_\alpha$, with $\alpha\in \{A,B\}$. 
This gives rise to a rich phase behavior and interesting dynamics.


\begin{figure*}
    \centering
    \includegraphics[width=\textwidth]{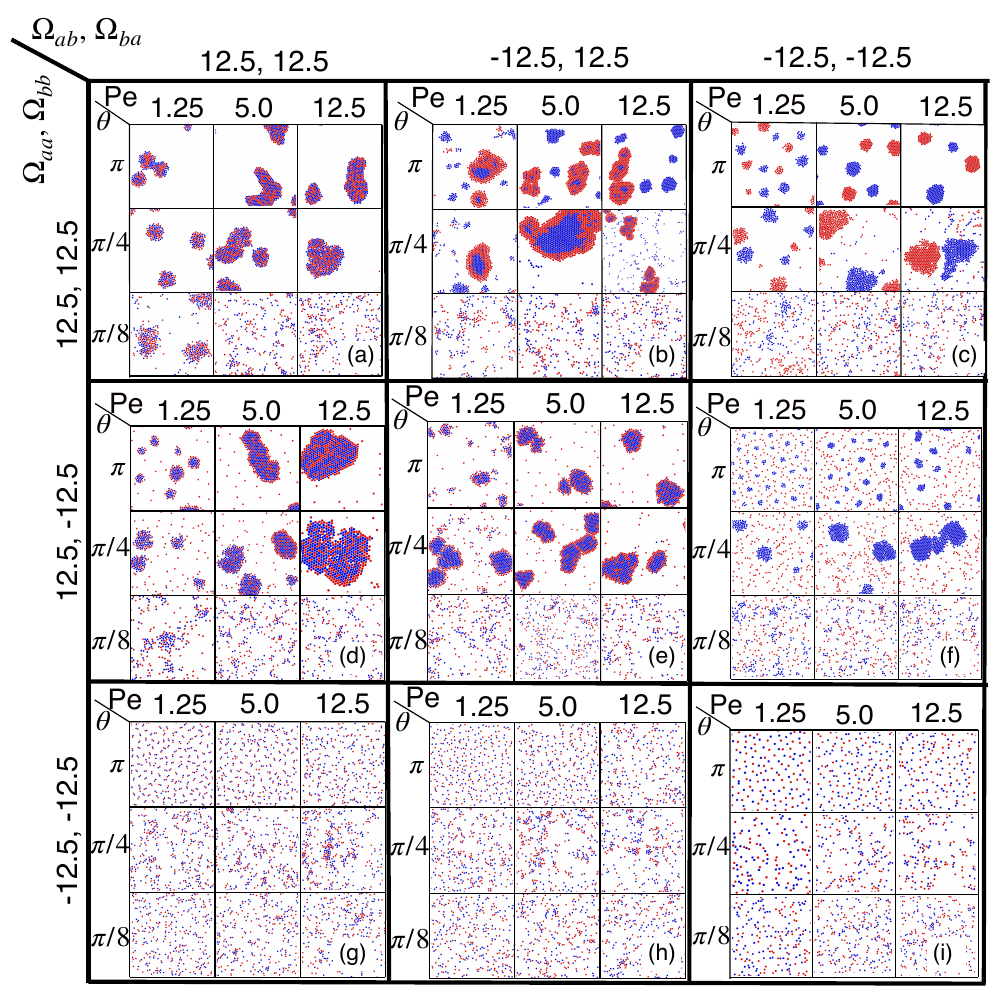}
    \caption{Snapshots of phases for $9$ combinations of the maneuverabilities and the Péclet numbers $1.25$,  $5.0$, 
    and $12.5$, and vision angles $\pi$,  $\pi/4$, and $\pi/8$. The number of particles is $N_A=N_B=N/2=500$ at  the 
    packing fraction $\Phi=0.0785$. (See also Movie M1 \cite{SI}.) } 
    \label{fig:global_phase}
\end{figure*}

\subsection{Parameters}

We measure lengths in units of the particle diameter $\sigma$, time in units of $\tau = \sqrt{m\sigma^2/(k_BT)}$, and energies in units of the thermal energy $k_BT$ \cite{negi_2022_soft_matter,negi_2024_PRR}. 

The friction coefficient $\gamma_T$ is chosen as $\gamma_T = 10^2 \sqrt{m k_BT/\sigma^2}$, the 
rotational diffusion coefficient as $D_R= 8 \times 10^{-2} \sqrt{k_BT/(m\sigma^2)}$, which yields the 
relation $D_T/(\sigma^2 D_R) = 1/8$ with $D_T = k_B T/\gamma_T$. 
These choices ensure that inertia effects are negligible, since the small ratio $m D_R/\gamma = 8 \times 10^{-4} \ll 1$
implies a strongly overdamped single-particle dynamics \cite{mandal_2019_PRL}. 
The activity of the iABPs is characterized by the P\'eclet number
\begin{equation}
     Pe=\frac{v_0}{\sigma D_R} . 
\end{equation} 
The adaptation of the interaction strength $\epsilon/k_BT=(1+Pe)$ warrants a nearly constant iABP overlap 
during iABP contacts, even at high activities \cite{negi_2024_PRR}. 
Periodic boundary conditions are applied.
The equations of motion \eqref{eq:gen_eom_pos} are integrated with a velocity-Verlet-based scheme suitable for 
stochastic systems \cite{gronbech_2013_MP}, using the time step $\Delta t = 10^{-3} \tau$. 
The maneuverabilities $\Omega_{\alpha \beta}$ are scaled with $D_{R}$, such that $\Omega_{aa}= \Omega_{AA}/ D_{R}$ etc. 
If not stated otherwise $R_{0}=1.5 \sigma$ and $R_{V} = 4 R_{0}$.
Initially, the iABPs are typically arranged on a square lattice, with iABPs distances equal to their 
diameter $\sigma$, in the center of the simulation box.

We performed $10^7$ relaxation steps, and collect data for the subsequent $10^7$ steps. For certain averages, 
up to $10$ independent realizations are considered.

\subsection{Phases and Phase Diagrams}
\label{sec:phasediagram_overview}

Figure \ref{fig:global_phase} presents an overview of emerging phases for nine combination of the 
maneuverabilities $\Omega_{\alpha \beta}$, with equal magnitude and all possible sign combinations. In each case, the 
three P\'eclet numbers $Pe=1.25, 5.0, 12.5$ and three vision angles $\theta_A=\theta_B = \theta = \pi, \pi/4, \pi/8$ 
are investigated. Various kinds of complex structures are obtained, such as mixed aggregates, segregated 
aggregates, layers of aggregates of one type of iABPs 
surrounded by a homogeneous layer of the other type of iABPs, homogeneous aggregates engulfed by the 
another type of iABPs, dimers, mixed states, and  honeycomb lattice type structures. 

In the following sections, the emerging structures and their dynamics will be discussed for several parameter 
combinations in more detail.

\section{Avoiding Same, Favoring Opposite Type of Intelligent ABPs} 
\label{BM:polymeric_aggregate}
\subsection{Structures}

\begin{figure*}
    \centering
    \includegraphics[width=.95\textwidth]{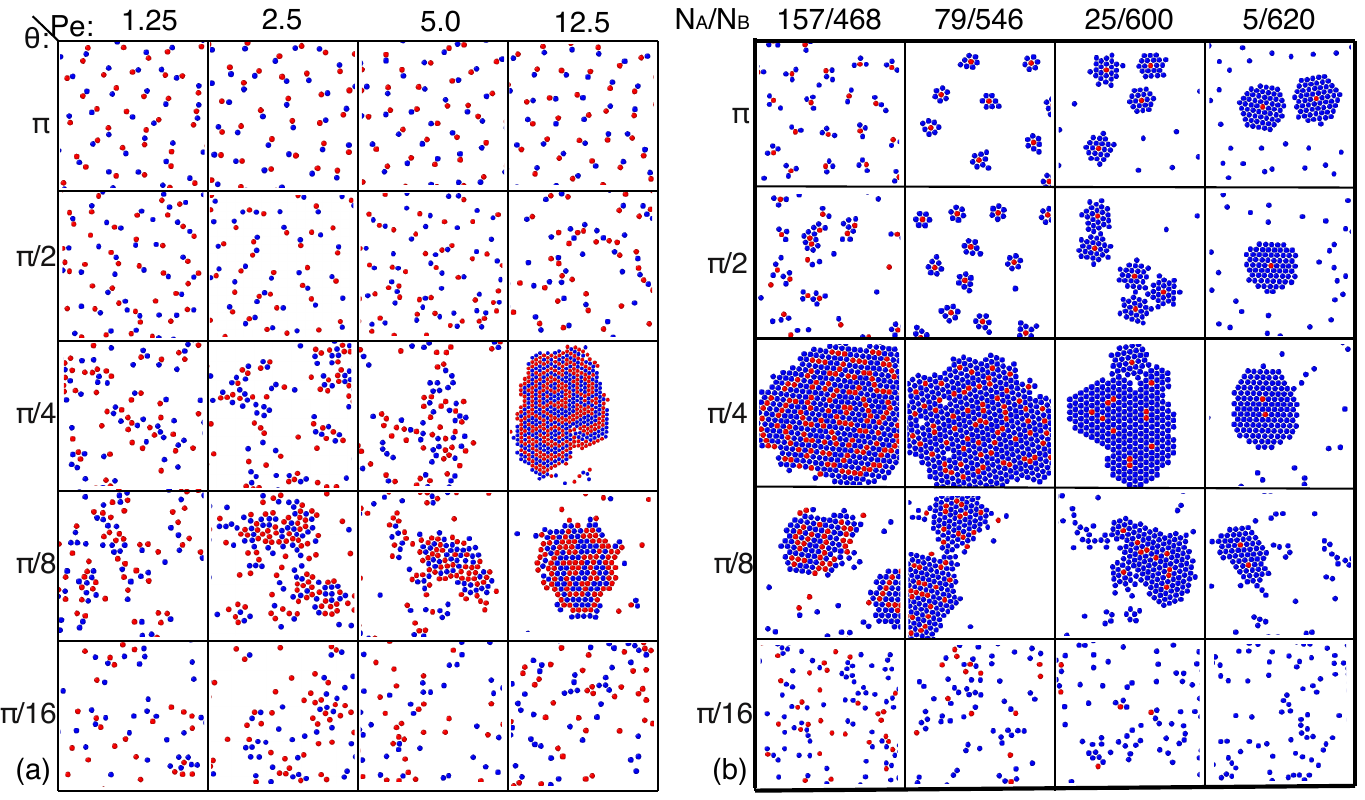}
    \caption{Snapshot of phases for the maneuverabilites $\Omega_{aa}=\Omega_{bb} =-62.5$ and $\Omega_{ab}= \Omega_{ba}=62.5$ and 
    various combinations of (a) vision angle $\theta_A=\theta_B = \theta$ and P\'eclet number $Pe$  for $N_{B}=N_{A}=312$, and 
    (b) and the ratio of the number $N_A/N_B$ of the two types of iABPs and the vision angle  for the P\'eclet number $Pe = 12.5$.  
    The packing fraction is $\Phi= 0.0785$. (See also Movie M2 and M3 \cite{SI}.)}  
    \label{fig:phase_diagram_across_ratios}
\end{figure*}

A system, in which particles of the same type steer away and of opposite type steer toward each other, 
i.e. $\Omega_{aa}=\Omega_{bb} <0$ and $\Omega_{ab}=\Omega_{ba} > 0$, bears 
some similarity with an electrostatic system, where equally charged particles repel and 
different charged particles attract. 
This analogy implies that the formation of ``charge" dipoles (dimers) can be expected. 
Snapshots of the typical conformations are displayed for $\Omega_{aa}=\Omega_{bb}= -\Omega_0$
and $\Omega_{ab}=\Omega_{ba}= +\Omega_0$ with $\Omega_0=12.5$ in 
Fig.~\ref{fig:global_phase}(g), and for $\Omega_0=62.5$ in Fig.~\ref{fig:phase_diagram_across_ratios}.
The emergent structures strongly depend on the vision angle and the P\'eclet number, as illustrated in 
Fig.~\ref{fig:phase_diagram_across_ratios}(a) for equal particle numbers $N_A=N_B$. 
This reflects the nonreciprocal character of the vision interaction. For the small maneuverability 
$\Omega_0 = 12.5$ and low $Pe$, Fig.~\ref{fig:global_phase}(g), as well as for 
$\Omega_0 = 62.5$ and all $Pe$, Fig.~\ref{fig:phase_diagram_across_ratios}, 
hetero-dimers form for the large vision angle $\theta = \pi$, as for electrostatic interactions. 
Nevertheless, no dimer pairs, chains, or clusters are found, as might be expected from the electrostatic analogy. 
This is a consequence of the polarity of the iABPs by their propulsion direction, which breaks 
the spatial isotropy of their mutual interactions, although vision itself is isotropic and 
stabilizes dimers, but leads to a repulsive interaction between 
them, so that the average distance between dimers is determined 
by the number $N_A=N_B$ of iABPs. For a homogeneous spatial distribution of particles, the average distance $d_p$
between dimers can be estimated from the density as
$d_p = (4/\pi)^{1/2}(L^2/N_A)^{1/2} \approx 5 \sigma$ (for $N=1000$ and $L =100$ as well 
as $N=625$ and $L=78.5$), in good agreement with 
simulation results. In contrast, for the small vision angle $\theta = \pi/16$, 
an iABP detects only other iABPs in a narrow vision cone -- the spatial symmetry of vision is broken -- and 
disordered structures of single particles are observed, especially for small $Pe$, as found in
non-steering systems.    

For vision angles $\pi/8$  and $\pi/4$,  larger P\'eclet numbers, and the higher maneuverability 
$|\Omega_{\alpha \alpha}|= |\Omega_{\alpha \beta}| = 62.5$, self-organized patters appear, see 
Fig.~\ref{fig:phase_diagram_across_ratios}(a). 
Even hexagonally close-packed structures are formed for large $Pe$. Specifically, double layers of 
similar iABPs reflect the nonreciprocal character of the interactions. 

It is of course also interesting to consider non-stochiometric systems, where one particle type,
say $A$, is the minority component, i.e.,  $N_A < N_B$. 
Figure~\ref{fig:phase_diagram_across_ratios}(b) illustrates the influence of the number ratio $N_A/N_B$ 
at constant overall packing fraction on structure formation. Instead of $A$-$B$ pairs, now clusters with a 
larger number of $B$-type iABPs appear. With increasing majority component $B$, 
clusters grow and several layers of $B$-type iABPs may form around $A$ particles. The cluster size is 
limited by (i) the number of minority iABPs, and (ii) the vision range.  For $\theta= \pi/4$ 
and $\pi/8$ and not too large ratios $N_B/N_A \approx 3$ and $7$, similar layered structures of $A$- and 
$B$-type particles appear as for the stochiometric case (at the same $Pe=12.5$), although with 
a smaller number on $A$-type iABPs.
For large vision angles $\theta\ge \pi/2$, the formation of triplets, septuplets, and larger aggregates is observed.

The (average) polarity field, determined by the orientation of the propulsion direction, in any cluster 
points preferentially toward a center of a cluster, although with some deviations 
in case of the opposite multilayers, see Fig.~\ref{fig:phase_diagram_across_ratios}(a). In particular, 
the clusters in Fig.~\ref{fig:phase_diagram_across_ratios}(b) for $N_A/N_B =620/5$ and $\theta = \pi$ with 
a single central $A$-particle are very stable (see Movie M1). The $B$-type particles all point toward the central $A$ 
particle, because the effective torque on a $B$ particle by its neighboring $B$ particles 
is close to zero, as (i) the torque on  head-to-tail arranged particles with parallel propulsion direction 
is zero (see Eq.~\eqref{eq:f_aa}), and (ii) the torque of the side-by-side arranged iABPs also nearly
vanished due to their hexagonal arrangement. That leaves a residual preferred motion toward the $A$
particle. Furthermore, the dispersed $B$ particles in the gas phase surrounding a cluster
repel those inside the cluster and thereby stabilize the phase separation.

\subsection{Dynamics}

\begin{figure}
    \centering
    \includegraphics[width=.45\textwidth]{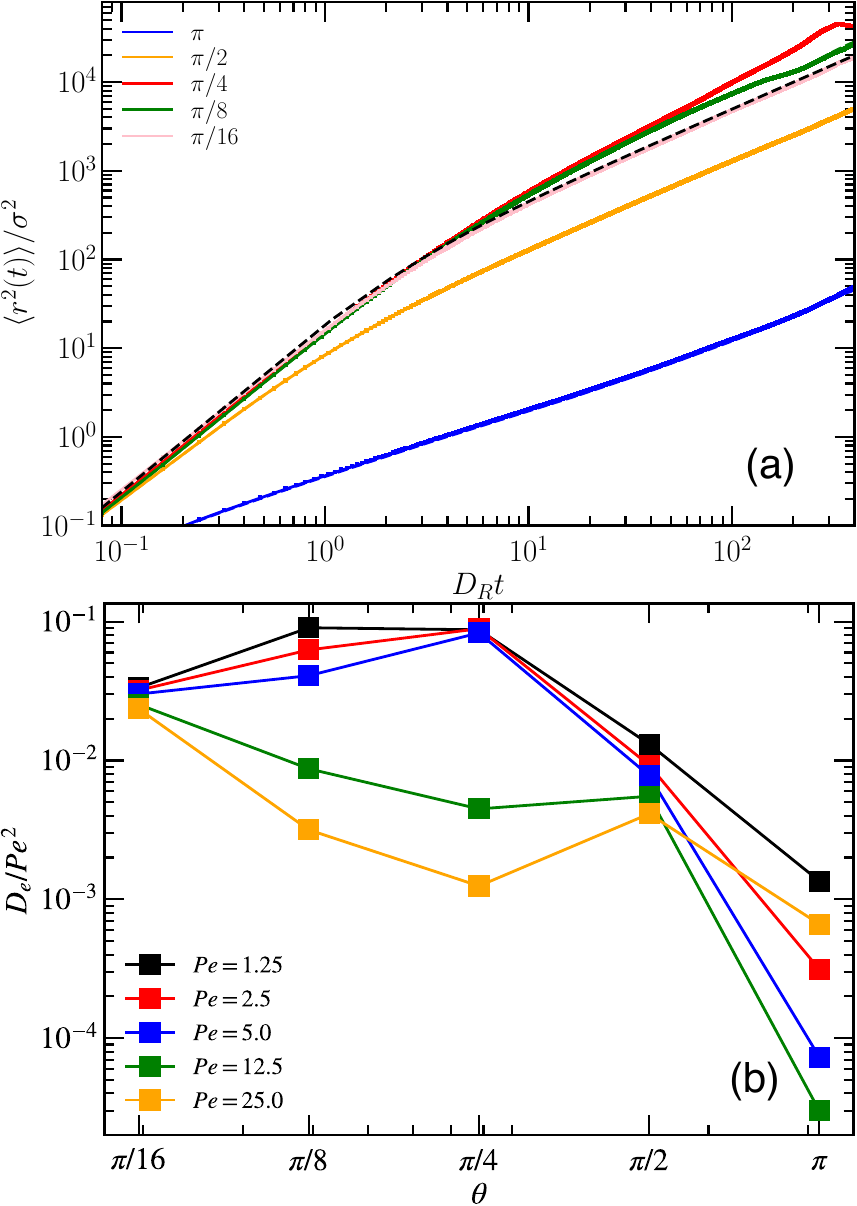}
    \caption{Dynamics of iABPs. (a) Mean squared displacement (MSD) as a function of scaled time $D_R t$ 
    for $Pe=5.0$ and various vision angles $\theta$.  The black dashed line represents the MSD of non-steering ABPs. (Will (a) be removed?)
    (b) Effective diffusion coefficient $D_e= D_{L}/D_R$ extracted from the MSD, scaled by the square of the P\'eclet number, as a function of the vision angle $\theta$ for various $Pe$. The number ratio is $N_{B}/N_{A}=312/313$, the packing fraction is $0.0785$.  }
    \label{fig:effectivemsd}
\end{figure}
To characterize the dynamical features of iABP self-organization, we consider their mean-square 
displacements (MSD) 
\begin{equation} \label{eq:gen_msd_eval}
    \langle \bm r^{2}(t) \rangle= \frac{1}{N} \sum_{i=1}^N \left\langle \left( \bm r_i(t+t_{0})- \bm r_i(t_{0}) \right)^2 \right\rangle ,
\end{equation}
where the average is performed over the initial time $t_0$. For vision angles $\theta \le \pi/2$, the dynamics shows an 
ABP-like behavior, with a ballistic motion and a diffusive motion for times $D_R t \gtrsim 10$, see 
Fig.~\ref{fig:effectivemsd}(a). To characterize the 
dynamics as a function of vision angle and P{\'e}clet number, we extract an effective translational diffusion 
coefficient $D_L$ from the long-time MSD, as displayed in Fig.~\ref{fig:effectivemsd}(b) as a function of the vision angle.

For large vision angle $\theta =\pi$, the MSD is very small due to the formation of dimeric clusters. 
The self-propulsion of dimers is ``screened'', because the propulsion directions point preferentially toward each other, as 
in a motility-induced phase separation of ABPs. In contrast to the latter, steering prevents a diffusive change of the 
propulsion direction and the pairs are stable for a long time. 

At smaller vision angle $\theta =\pi/2$, dimers are less stable and the formed structures are more disordered, also 
larger temporary string-like (polymer-like) aggregates appear. The corresponding MSD is larger than that for $\theta=\pi$, 
but smaller than that for a non-steering ABP due to the formation of small clusters.  Interestingly, an enhanced diffusion 
is also observed for the vision angles $\theta = \pi/4$ and $\pi/8$, which can be attributed to temporarily formed small 
clusters with highly inhomogeneous particle orientations, where the restricted rotational diffusion due to vision 
interaction causes a slightly enhanced persistent motion.  Finally, at the very small vision 
angle  $\theta = \pi/16$, where an iABP detects hardly any other iABPs, the system is in a disordered dilute phase, 
where the particles exhibit the same MSD as individual non-steering ABPs.

This behavior is of course also reflected in the results for the diffusion coefficient $D_e$. It first increases with 
increasing $\theta$, then decreases for $\theta \gtrsim \pi/4$, as long as $Pe \lesssim  5.0$. 
In this range of $Pe$ values, $D_e$ is nearly independent of $Pe$, because motion is dominated by the translational 
diffusion of dimers. For larger P\'eclet numbers, $Pe=12.5$ and $25.0$, large aggregates are present for 
$\theta = \pi/4$ and $\pi/8$ (compare Fig.~\ref{fig:phase_diagram_across_ratios}(a)), where the total propulsion is 
reduced by partial compensation of the 
particle activity. The $N_e$ ``passivated" particles in the cluster diffuse with an effective diffusion coefficient 
$\sim 1/N_e$, which reduces the overall diffusion coefficient. Interestingly, $D_e$ is larger for $Pe \leq 5.0$ and 
$\theta \le \pi/2$ than for larger $Pe$, in particular for $Pe=25.0$. This happens, because higher $Pe$ stabilizes 
propulsion directions toward the cluster center. Finally, for $\theta=\pi/16$, the diffusion 
coefficients are well fitted by the relation $D_e = D_{L}/D_R \approx Pe^{2} = v_{0}^2 \sigma^2/(D_{R}^2 ) $ of non-steering ABPs. 

The clusters in Fig.~\ref{fig:phase_diagram_across_ratios}(a) for $\theta = \pi/4$ and $Pe=12.5$, and those in  Fig.~\ref{fig:phase_diagram_across_ratios}(b) for $\theta = \pi$ and $N_A/N_B = 620/5$ exhibit an interesting 
collective dynamics. In the first case, a pronounced rotational motion appears due to a polarization of the 
propulsion directions. In the second case, the the near-hexagonal arrangement of the $B$-type particles 
leads to a compensation of propulsion forces, and the cluster translates only very slowly with very 
little rotation.

\begin{figure}
    \centering
    \includegraphics[width=.5\textwidth]{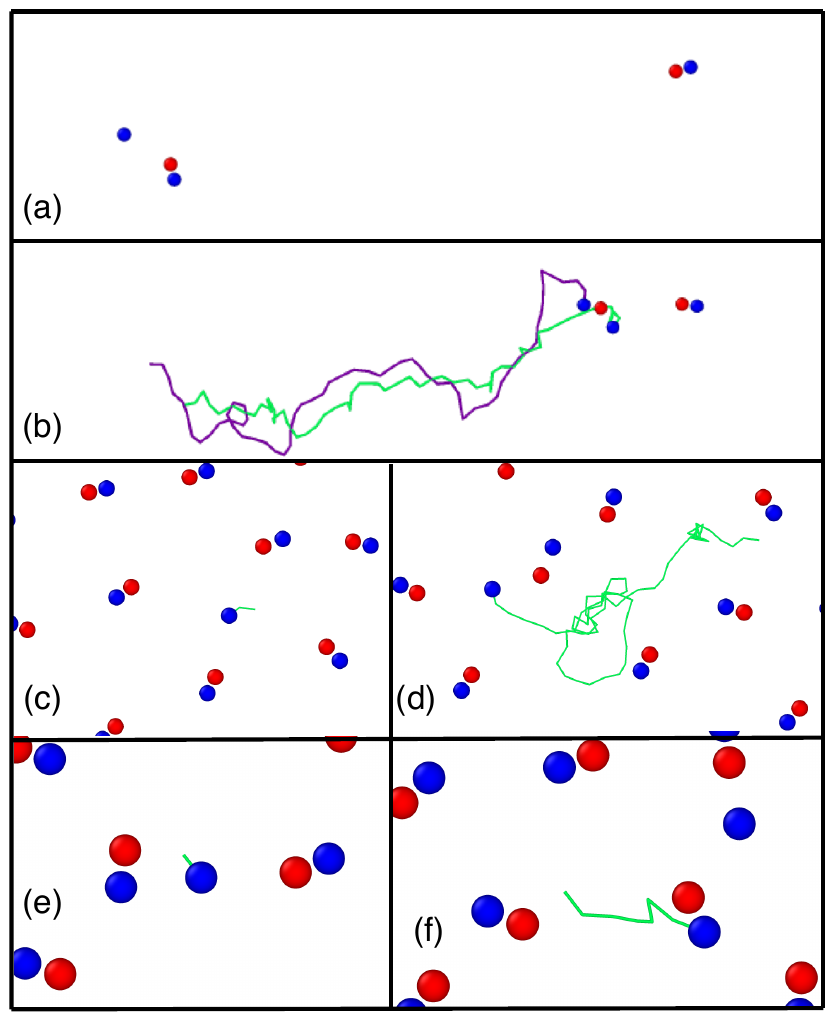}
    \caption{Snapshots of the hopper-pair exchange dynamics. The depicted sequences illustrate the replacement of a 
    particle within a pair by a ``hopper'', leading to the formation of a new hopper. 
    Initially (a), (c), (e) a blue  hopper, steers toward the red particle of a pair at the packing fractions 
    (a), (b) $\Phi =0.00785$, (c), (d) $\Phi =0.03925$, and (e), (f) $\Phi =0.0785$. 
    (b), (d), (f) Subsequently to an exchange, a blue hopper traverse the green trajectory and substitutes another 
    blue particle within the pair. The parameters used are $N_{d}=5$ and $Pe=12.5$. (See Movie M4 \cite{SI}.)}
    \label{fig:hopping_traj}
\end{figure}

\subsection{Hopping}
\label{Hopper}
In the nearly stochiometric case, with $N_B = N_A + N_d$ and $N_d \ll N_A$, $A$-$B$ dimers dominate in the 
conformations but are accompanied by $N_d$ unpaired iABPs. These unpaired particles can shuttle between dimers 
and exchange places with particles within the dimers. Hence, we  denote such an unpaired particle 
as a hopper in the following. The transport of unpaired particles resembles the Grotthuss mechanism of proton 
transport from one water molecule to another via the formation and breaking of hydrogen bonds \cite{agmon_1995_grotthuss}. 
However, the transport of hoppers between dimers is by active motion and not by hydrogen bond rearrangements 
along a network, as in the Grotthuss mechanism \cite{agmon_1995_grotthuss}.

Figure \ref{fig:hopping_traj} shows sequences of such hopping events, where a particle from a pair is replaced by a 
hopper, resulting in the formation of a new pair. To define a hopper, we use a distance criterion -- as long as a particle 
is not within the distance $h_{d}= 1.5 \sigma$ of another particle, it is a hopper. 
The dynamics of hoppers depends on the  P\'eclet number, the packing fraction, and the number 
$N_{d}= |N_{B}- N_{A}|$ of excess $B$- or $A$-type particles beyond stochiometry.

\subsubsection{Hopper number, encounter distance, displacement, and hopping time}
\begin{figure}
    \centering
    \includegraphics[width=\columnwidth]{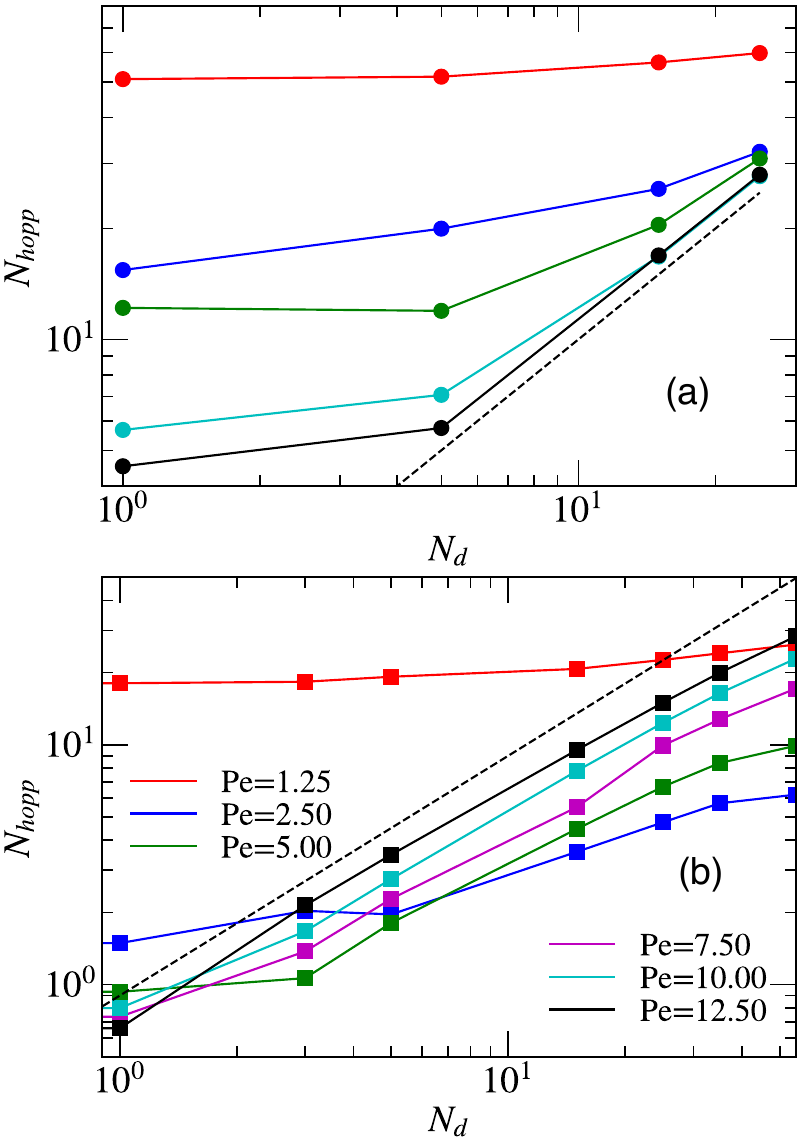}
    \caption{Activity dependence of the number of hoppers. Number $N_{hopp}$ of hoppers as a function of the number difference 
    $N_{d}= N_{B}-N_{A}$ for various P\'eclet numbers ($Pe$) and the packing fractions (a) $\Phi=0.00785$ and (b)  $\Phi=0.0785$. The dashed lines correspond to $N_{hopp}=N_{d}$.} 
    \label{effective_n_h}
\end{figure}

Figures \ref{effective_n_h}(a) and (b) show the average number $N_{hopp}$ of hoppers present in the system at 
various activities and packing fractions. Interestingly, at small $Pe=1.25$ the number of hoppers is larger compared to 
higher $Pe$, and increases only slightly as the number difference $N_{d}$ increases. This larger number $N_{hopp}$ is due 
to thermal noise. Despite opposite-type particles steering toward each other, noise disrupts their sustained proximity,
and suppresses long-time stable pair formation, which results in the creation of many hoppers. At the low density $\Phi = 0.00785$ (Fig.~\ref{effective_n_h}(a)), the number of hoppers decreases with increasing activity, as dimers become more stable.  
$N_{hopp}$ 
decreases with increasing packing fraction at fixed $Pe$ and $N_d$, as the distance between dimers decreases
and they become less susceptible to a particle exchange (see Figs.~\ref{effective_n_h}(a) and (b)). For the higher packing fraction $\Phi=0.0785$ (Fig.~\ref{effective_n_h}(b)), 
$N_{hopp}$ is smaller than for the lower packing fraction (Fig.~\ref{effective_n_h}(a))  (at a given $Pe$ and $N_d$). 
Due to the larger density, hoppers
are more likely to meet dimers, but the exchange process takes longer, which results in more stable clusters of three 
particles and, hence, a reduced number of hoppers. At large $Pe$ and $N_d$ -- the latter number depends on the packing 
fraction -- $N_{hopp}$ increases approximately linearly with the number difference, which suggest that approximately all 
minority-type $A$ particles are bound in pairs. 

\begin{figure}
    \centering
    \includegraphics[width=\columnwidth]{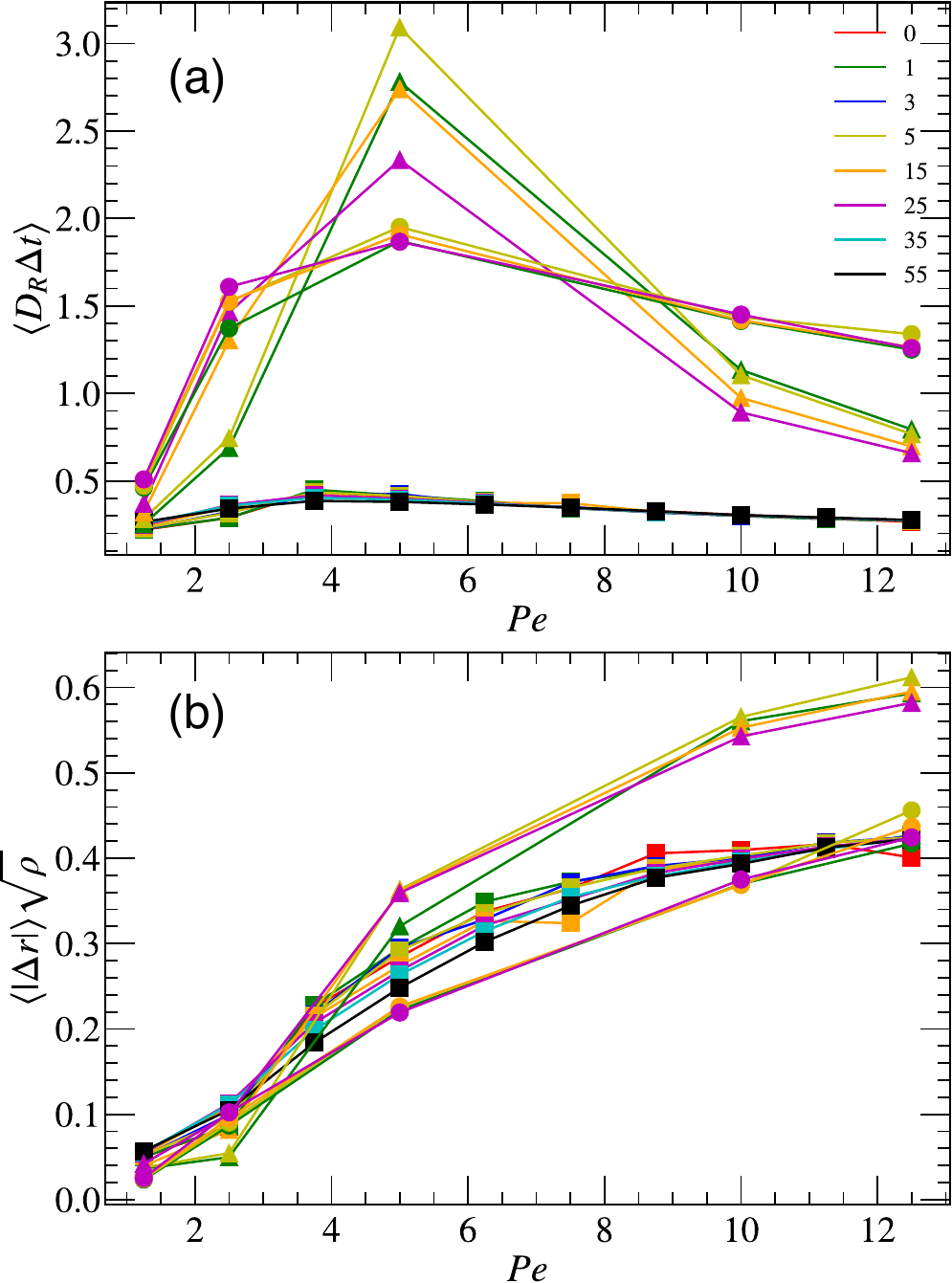}
    \caption{Characteristic average life times and displacements of hoppers as a function of P\'eclet number and for various number differences $N_d$ (legend). (a) Average hopping time $\langle D_R \Delta t \rangle$, during which a hopper is unbound before combining with other particles, and
    (b) average scaled displacement  $\langle |\Delta \bm r| \rangle \sqrt{\rho}$
    as a function of the P\'eclet number $Pe$ for various number difference $N_{d}$ (see legend). 
    Here, $\rho=(N_A+N_B)/L^2$ is the number density of iABPs.   
    Packing fractions are indicated by different symbols, with $\Phi= 0.0785$ (squares), $0.03925$ (triangles), and 
    $0.00785$ (bullets).}
    \label{mean}
\end{figure}

To further characterize the hopping behavior in the process of the formation of a new pair, we calculate the average 
hopping time $\Delta t$ and the average hopping displacement $|\Delta \bm r|$.  The results presented in 
Fig.~\ref{mean} reflect a distinct density dependence. For higher packing fractions, 
hoppers traverse small distance only and thus have a high encounter probability with $A$-$B$ pairs or other free particles,
which expedites the of hopper-transfer process. 
Conversely, for lower packing fraction, the hopping times increase
similar to the increasing traversed distances. The shortest average hopping time occurs at P\'eclet number $Pe=1.25$, 
again due to the dominance of noise. As particles in a pair are loosely attached to each other, hoppers can easily 
break a pair and generate a new pair and a new hopper. As the activity increases ($Pe \lesssim 5.0$), dimers are 
more stable and the hopping time increases. For even larger $Pe$, $D_R \Delta t$ is almost constant or decreases 
with in increasing $Pe$, depending on density, due to the enhanced persistent motion of the iABPs. 


The simulation results for the hopping time $\Delta t$ suggest that it is mainly determined by the distance a hopper covers 
before exchanging with another pair, see Fig.~\ref{mean}(a). This is different for the displacement $|\Delta \bm r|$ 
displayed in Fig.~\ref{mean}(b), which scales very well with the average dimer distance $1/ \sqrt{\rho}$, and increases 
monotonically with increasing P\'eclet number and saturates for large $Pe$. 
Notably, $|\Delta \bm r|$ is very similar for the largest and smallest considered packing fraction. Moreover, 
the two quantities presented in Figs~\ref{mean}(a) and (b) are independent of the number difference 
for packing fraction $\Phi=0.0785$ and $\Phi=0.00785$.   

Remarkably, the systems at the intermediate packing fraction $\Phi=0.03925$ exhibit a different behavior. First, 
the average hopping time depends on the number difference. Second, the maximum average hopping time is larger 
than the values for the larger and smaller packing fractions. We conjecture that this is related to 
collective effects. At this intermediate density, hoppers are typically equidistant from other pairs, hence 
selection and motion toward the closest dimer is ambiguous. In addition, pairs are more responsive to an approaching 
hopper compared to higher density systems. Thus, a hopper requires a longer time to reach a dimer and replace a 
paired iABP.

\subsubsection{Mean-Square Displacement: Caging and Chasing}
\begin{figure}
    \centering
    \includegraphics[width=.48\textwidth]{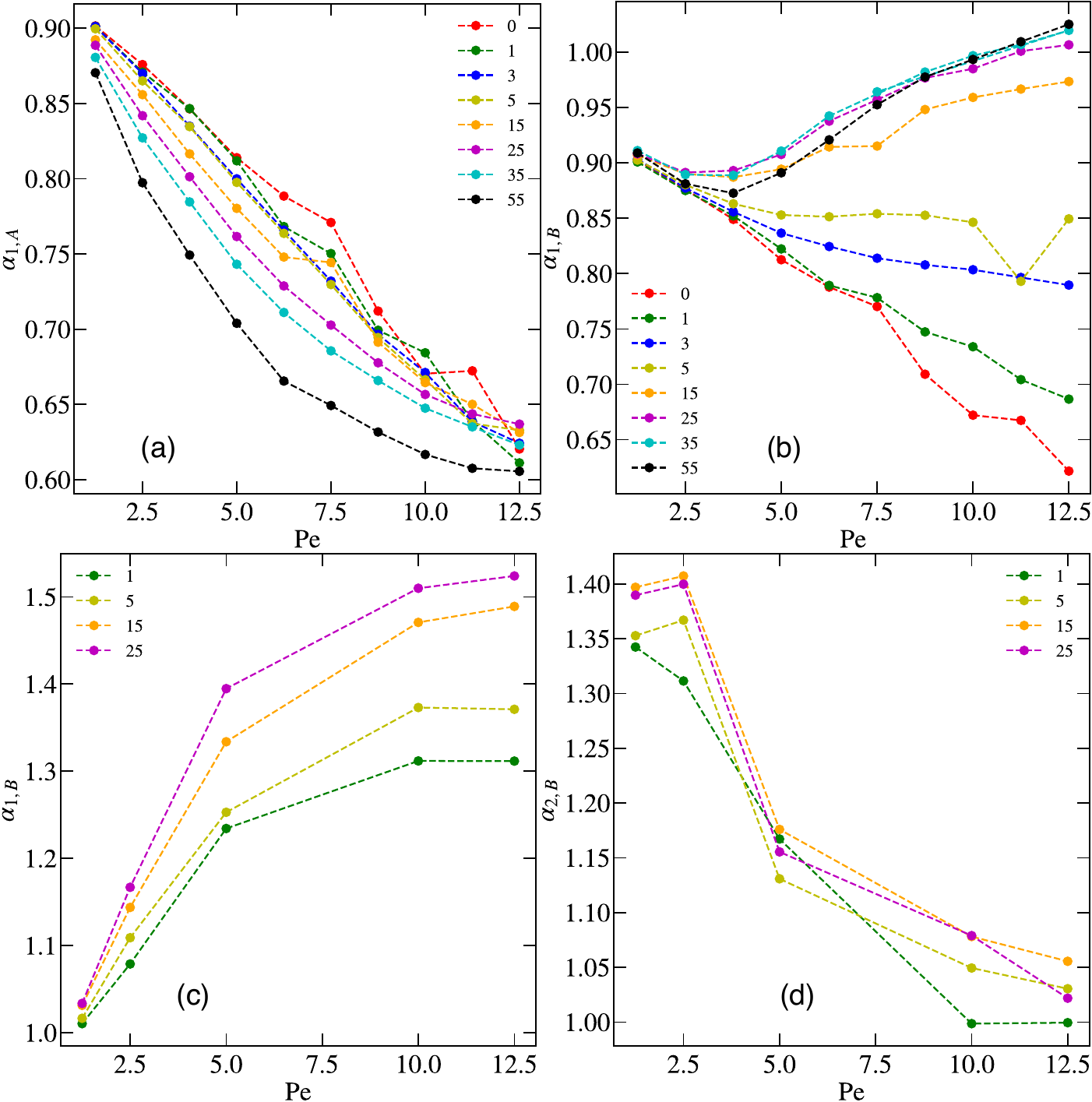}
    \caption{Dynamics of $A$- and $B$-type iABPs, characterized by the MSD exponent
    $MSD \sim t^\alpha$ for the short and intermediate time range. 
    (a) Exponents $\alpha_{1,A}$ for $A$ (minority)  particles and  
    (b) $\alpha_{1,B}$ for $B$ (majority) particles at the packing fraction $\Phi=0.0785$; 
    (c) exponents $\alpha_{1,B}$ and 
    (d) $\alpha_{2,B}$ for $B$ particles as a function of the P\'eclet number $Pe$ and various $N_{d}$ 
    (legend). The total number of particles is $N=N_A+N_B=625$.
    }
    \label{fig:msd_hoppers}
\end{figure}

To characterize the dynamics of iABPs in the ''hopping phase'', we evaluate the mean-square displacement (MSD) 
of the $A$ and $B$ particles. The characterization of the dynamics of hoppers itself is 
only possible for rather short times due to frequent recombination and exchange with $A$-$B$ pairs.  
The presence of the majority $B$ component affects the dynamics of the mixture. Depending 
on the parameters, we find short- and intermediate-time regimes with sub- and super-diffusive behavior -- the 
long-time dynamics (with $D_R t \gtrsim 40$) is always diffusive. 
The various MSD regimes are differentiated by their power-law time dependent MSD $t^{\alpha_{1,2}}$, with exponents 
$\alpha_1$ in the short-time ($D_R t < 2$) and $\alpha_2$ in the intermediate-time ($2< D_R t < 20$) regime.

Figures \ref{fig:msd_hoppers}(a) and (b) show values of the exponent $\alpha_{1,A}$ of $A$ (minority) and $\alpha_{1,B}$ of 
$B$ (majority) particles for the large packing fraction $\Phi=0.0785$. The values $\alpha_{1,A} <1$ (see Fig.~\ref{fig:msd_hoppers}(a)) 
reflect a subdiffusive nature of the minority $A$ particles, which are typically 
paired with $B$ particles, which repel other $B$ particles and thus exhibit a caging effect (see Figs.~\ref{fig:global_phase}(g), 
\ref{fig:phase_diagram_across_ratios}(b)). The exponent $\alpha_{1,A}$ decreases with increasing activity as the propulsion 
directions in these dimeric pairs are more strongly aligned toward each other, which reduces the overall (active) dynamics. 
The weak dependency on  $N_{d}$, with the lowest value of $\alpha_{1,A}$ at large $N_{d}=55$ and the highest at low $N_{d}=1$, 
can be attributed to the fact that the larger the number difference the smaller is the long-time stability and mobility 
of individual dimers.

The majority $B$-type iABPs exhibit nearly diffusive behavior for $N_{d} \lesssim 5$ and $Pe < 2.5$ 
(see Fig.~\ref{fig:msd_hoppers}(b)). However, the exponent $\alpha_{1,B}$ decreases with increasing $Pe$. This is a 
consequence of the more stable $A$-$B$ pairs at larger $Pe$, which leads to a drop in the number of (highly mobile)
hoppers (see Fig.~\ref{effective_n_h}). The $A$ and $B$ particles exhibit 
then a similar, cage-determined diffusive behavior (cf. Fig.~\ref{fig:msd_hoppers}(b)). With increasing $N_d$, more free 
hoppers are present (Fig.~\ref{effective_n_h}) and $\alpha_{1,B}$ increases, approaching the diffusion limit $\alpha_{1,B}=1$ 
for $N_d \gtrsim 15$, nearly independent of the P\'eclet number.

The dynamics of the iABPs is rather different in the dilute systems with packing fraction $\Phi=0.00785$. The minority $A$ 
particles are as mobile 
as the $B$ particles, consequently the values of the exponents are $\alpha_{1,B} \approx \alpha_{1,A}$ and 
$\alpha_{2,B} \approx \alpha_{2,A}$. In fact, the MSD is smaller for $B$s than for $A$s, because the opposing 
propulsion directions in pairs reduce the activity contribution to their dynamics. Propulsion implies a super-diffusive 
dynamics at short and intermediate time scales, with exponents $\alpha_{1,B} >1$ (for $Pe>2.5$) and $\alpha_{2,B} >1$ 
(for $Pe<10.0$), where an increasing activity leads to an increasing exponent at short times (see Fig.~\ref{fig:msd_hoppers}(c)) 
and a decreasing exponent for intermediate times (see Fig.~\ref{fig:msd_hoppers}(d)). The growth of the exponent 
$\alpha_{1,B}$ can be attributed to an enhanced 
mobility of the iABPs by the increasing activity, in analogy to that of active Brownian particles. Interestingly, hoppers can 
chase $A$-$B$ pairs (compare Fig.~\ref{fig:hopping_traj}), in which the propulsion directions turn approximately in the same direction, hence, leads to an enhanced 
dynamics of the hopper and the pair. As $N_{d}$ increases, the number of hoppers which chase a pair increases, thus, the 
effective mobility of both increases too, which is reflected in a growth of the exponent $\alpha_{1,B}$ with increasing $N_{d}$. 
In the intermediate time regime, $2< D_R t < 20$, hoppers are super-diffusive at small P\'eclet numbers. This is 
explained by an activity-enhanced dynamics as in the short-time regime for large $Pe$. With increasing $Pe$, the activity-enhanced 
time regime shifts to shorter times, and in the intermediate regime hoppers exhibit the long-time diffusive behavior, which 
is only weakly dependent on the number difference $N_d$.       


\section{Predator-Prey Behavior} 
\label{sec: Prey_Predator}

\begin{figure*}
    \centering
    \includegraphics[width=0.9\textwidth]{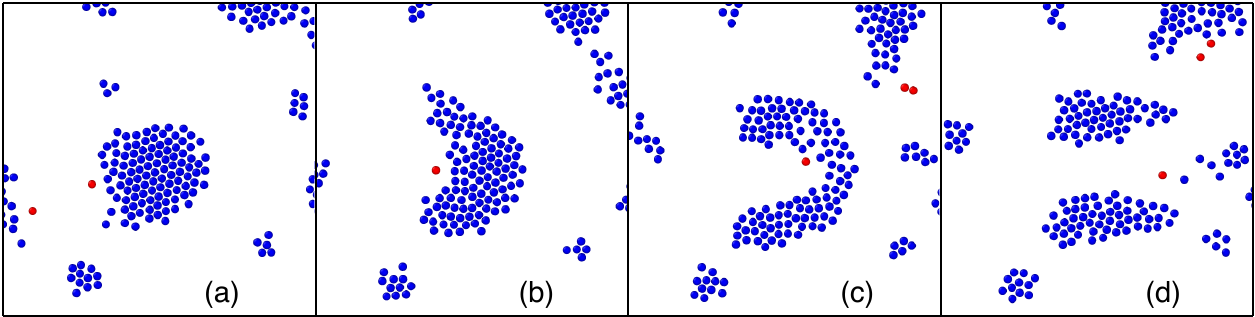}
    \caption{Snapshots of a predator-prey pursuit. Series of snapshots (zoomed) progressing in time from (a)-(d), 
    which illustrate a typical pursuit of a predator (red) chasing prey particles (blue). The parameters are 
    $N_{B}/N_{A}=50/1$, $\theta_{A}=\pi/4$, $\theta_{B}= \pi$, $\Omega_{aa}=\Omega_{bb}=12.5$, 
    $\Omega_{ab}=12.5$, $\Omega_{ba}=-12.5$, and the number of particles $N_A+N_B =1000$. 
    (See also Movie M5 \cite{SI}.) }
    \label{fig:predator_prey}
\end{figure*}

When like-particles attract each other, while $A$ particles chase $B$ particles, which try to 
escape from $A$ particles -- with maneuverabilities $\Omega_{aa}=\Omega_{bb} = + \Omega_0$, and 
$\Omega_{ab} = - \Omega_{ba} = \Omega_0$ -- a predator-prey-type of collective behavior is observed. 
This combination of the signs of maneuverabilites leads to the formation of cohesive groups and clusters by the 
same kind of particles (compare Fig.~\ref{fig:global_phase}(b)).
We consider here a system, where the vision angles of the two particle types are {\em not} the same, but 
a narrow vision angle $\theta_A=\pi/4$ for the predator, and a large vision angle $\theta_B=\pi$ for the prey. Note that throughout this simulation study we consider the case where the P{\'e}clet number is the same for $A$ and $B$ particles, so that the speeds of predator and prey are the same!
The emerging clusters can be considered as ``hunting packs'' of type-$A$ 
particles chasing ''herds'' of prey-like type-$B$ particles. Figure~\ref{fig:predator_prey} displays a sequence 
of snapshots illustrating a typical pursuit scenario (see also Movie M5). Clearly, the prey particles 
form various coexisting 
clusters, while the number of predator particles is too small for the formation of larger clusters. 
As time moves on, the predator approaches a prey cluster, and the prey particles try to escape, while the predator 
maintains its moving direction until the prey cluster has dispersed and the predator encounters another prey cluster.

%
\begin{figure}
    \centering
    \includegraphics[width =\columnwidth]{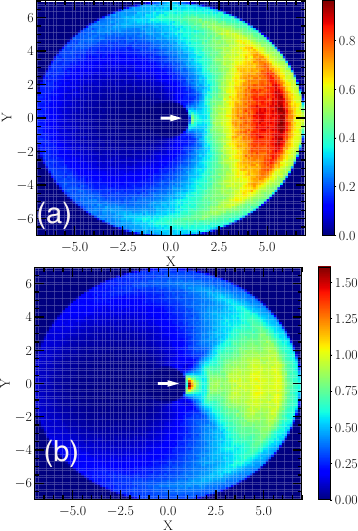}
    \caption{Distribution of the number of prey particles around a predator for the  maneuverabilities  (a) $\Omega_{ab}=12.5$ and (b) $\Omega_{ab}=50$. 
    The white arrow in the center indicates the predator's moving direction. 
    The other parameters are $\Omega_{aa}=\Omega_{bb}=12.5$, $\Omega_{ba}=-12.5$, $Pe=1.25$, $\theta_{A}=\pi/4$, 
    $\theta_B=\pi$,  $N_{B}/N_{A}=50/1$, and $N_A+N_B = 1000$. }
    \label{area}
\end{figure}
The spatial distribution of prey particles around a predator is displayed Fig.~\ref{area} for various maneuverabilities 
$|\Omega_{ab}|$, averaged over several encounters of predator and prey. The distribution of prey is symmetric in 
the direction perpendicular to the predator's moving direction and asymmetric along the propulsion direction, 
with a depletion of the number of prey particles behind the predator. As the prey senses the predator, it steers 
away from it. Since the small vision angle of the 
predator, $\theta_A=\pi/4$, provides only a limited view on the prey and its steering maneuverability, $|\Omega_{ab}|=12.5$, 
is low, the predator moves toward the prey in front, and most of the prey 
can escape sidewise and keep at a reasonable distance of approximately
$4.5 \sigma$ from the predator. However, when the predator's steering maneuverability is increased to $|\Omega_{ab}|=100$, 
the predator can catch up with the prey more easily, and move closer to 
the prey.

\begin{figure}
    \centering
    \includegraphics[width= \columnwidth]{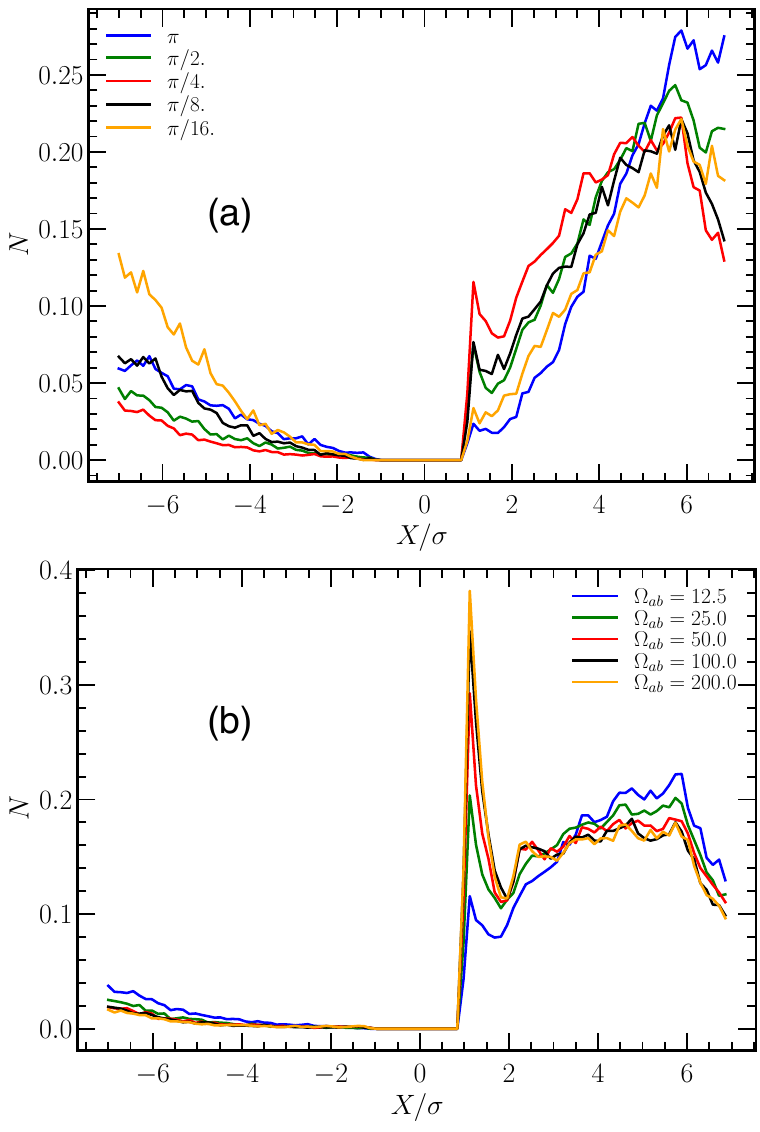}
    \caption{Number of prey particles ($B$) in the direction of motion of the predator ($A$) for (a) various  vision angles $\theta_{A}$ and $\Omega_{ab}=12.5$, and (b) various maneuverabilities and $\theta_{A}=\pi/4$.  The other parameters are the same as in Fig.~\ref{fig:predator_prey}.}
    \label{fig:number_dis}
\end{figure}

A quantitative characterization of the prey distribution in the moving direction of the predator is shown in Fig.~\ref{fig:number_dis} for various vision angles $\theta_A$ (Fig.~\ref{fig:number_dis}(a)) and maneuverabilities $|\Omega_{ab}|$ (Fig.~\ref{fig:number_dis}(b)). As a predator moves closer to a group of prey, prey particles sensing the predator disperse first by moving sidewise and reconvene later. Due to the cohesion of the prey cluster, even prey particles outside of the vision range of the predator move away from it and become depleted in its vicinity. Hence, the number of prey particles is drastically reduced behind the predator ($x\leq 0$). Conversely, in front of the predator, there is a substantial accumulation of prey that the predator is actively steering toward. For smaller predator vision angles, such as $\pi/16$, the visual field is narrower, 
leading to a reduced prey detection and subsequently lower prey densities. As $\theta_A$ increases to $\pi/8$ and 
$\pi/4$, the prey density reaches its maximum in front of the predator. A further increase of $\theta_A$ to $\pi/2$ 
and $\pi$ leads again to a decline of the prey density. Hence, the optimal vision angle of a predator to hunt prey
lies around $\pi/4$, which provides focused vision (``eagle's eye"). The broader field of view for 
$\theta_A \ge \pi/2$ counteracts effective steering toward prey, because the visibility of prey in 
many different directions implies a less goal-oriented and more erratic 
motion of the predator. 

Figure \ref{fig:number_dis}(b) illustrates the prey's number distribution around the predator for the angle 
$\theta_{A}=\pi/4$ and various maneuverabilities $\Omega_{ab}$. Again, an asymmetric distribution evolves, which is characterized by 
the scarcity of prey behind the predator and an accumulation in front. At large predator maneuverability, 
$|\Omega_{ab}|=200$, the probability for the predator to encounter a large number of prey particles 
in close proximity is high, because the predator can rapidly adjust its orientation and direction of movement, optimizing its chance of closely approaching and catching  the prey.
This probability diminishes as the maneuverability $|\Omega_{ab}|$ decreases, as the predator struggles to 
efficiently position itself in relation to the prey, resulting in a reduced likelihood of the 
predator to closely approach the prey.


\section{Honeycomb-Lattice-type Structure Formation}
\begin{figure}
    \centering
    \includegraphics[width=\columnwidth]{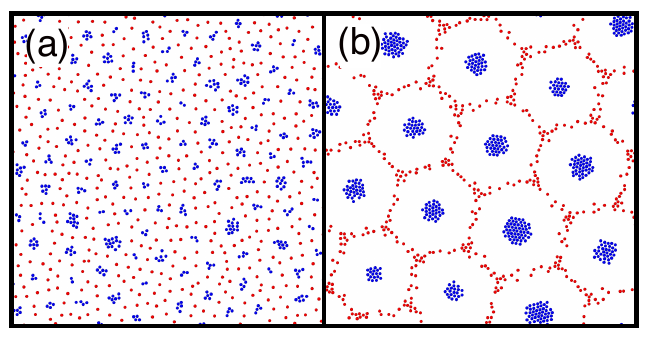}
    \caption{Snapshots of honeycomb-type lattices. The various structures follow by varying the vision range $R_v$, 
    where in (a) $R_v = 4.0 \sigma$, and (b) $R_v = 14.0\sigma$. 
    The maneuverabilities are $\Omega_{aa}=-12.5$, $\Omega_{aa}=12.5$, $\Omega_{ab}=\Omega_{ba}= -12.5$. 
    Furthermore, $Pe =1.25$, $N_A+N_B=1000$, $\theta_A=\theta_B =\pi$, and $\Phi = 0.0785$. 
    (See also Movie M6 \cite{SI}.) }
    \label{fig:snapshot_honecomb}
\end{figure}
We want to illustrate the enormous variability of self-organization in mixtures of self-steering active
agents by briefly discussing another intriguing phase, a quasi-periodic lattice structure, honeycomb-like lattice. Such a phase is obtained when one particle
type ($B$) wants to aggregate, the other ($A$) to disperse, while different types
want to avoid each other.
This happens for the maneuverabilities $\Omega_{aa}=-\Omega_{bb}<0$, and $\Omega_{ab}=\Omega_{ba}<0$. Figure \ref{fig:snapshot_honecomb} 
displays snapshots for $N_A=N_B$, which reflect the dependence of the emerging lattice 
on the vision cutoff range $R_v$ (see also Fig.~\ref{fig:global_phase}(f)). When the vision cutoff 
range is small (Fig.~\ref{fig:snapshot_honecomb}(a), $R_v=4.0\sigma$), particle segregation is weak, only
small clusters from and locally diffuse honeycomb-like structures occur. As the vision range increases to $R_{v} = 14.0\sigma$ 
(Fig.~\ref{fig:snapshot_honecomb}(b)), a pronounced and well-defined honeycomb lattice is formed. This extension of the cut-off 
range increases the number of particles in specific areas, the lattice structure becomes better defined, and a 
noticeable number of particles cluster together in the center of each hexagon as well as along its edges. Thus, 
the visual interaction range can play a crucial role in the formation of particular structures of iABPs. 

\begin{figure}
    \centering
    \includegraphics[width=\columnwidth]{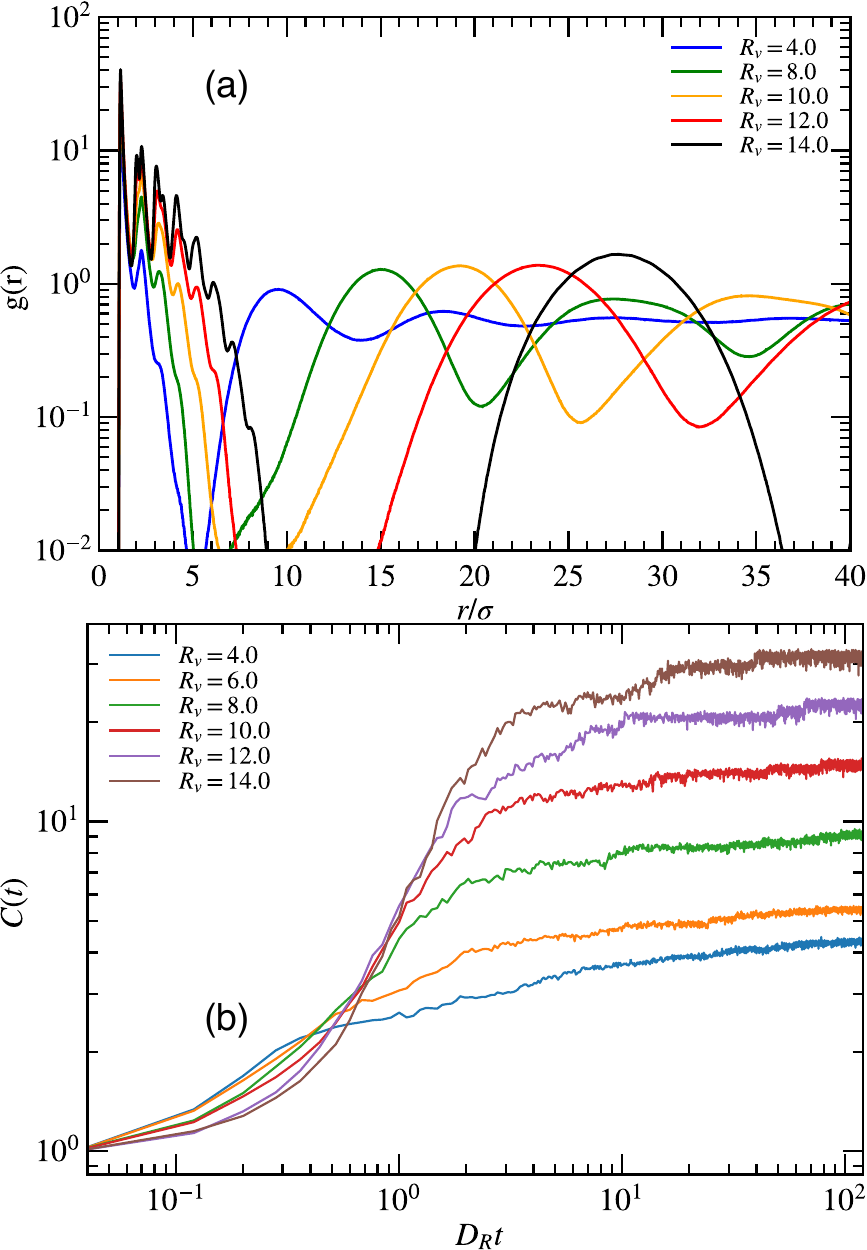}
    \caption{
    (a) Pair correlation-function of $B$ particles as a function of the radial distance for various vision cutoff radii $R_v$ (legend).  
     (b) Average number $C(t)$ of $B$ iABPs in a cluster as a function of the scaled time $D_R \Delta t$ for various 
     vision cutoff radii $R_v$. The P\'eclet number is $Pe=1.25$. The other parameters are the same as in Fig.~\ref{fig:snapshot_honecomb}.
     }
    \label{fig:binary_cluster}
\end{figure}

Figure~\ref{fig:binary_cluster}(a) displays the pair correlation-function for $B$ particles, which 
demonstrates the prevalence of the honeycomb lattice structures. At small $R_v$, small cluster of 
$B$ particles are formed and a fluid-like distribution appears at distances $r/\sigma >5$. With 
increasing $R_v$, the core clusters grows with pronounced peaks at the various $B$ particle layers. 
In addition, a broad peak grows and shifts to larger distances with increasing $R_v$, indicating the 
position of the first hexagonal shell of neighboring $B$ particle clusters, where its width accounts 
for the average number of $B$ particles in the respective clusters.

It is also interesting to consider the dynamics of the structure formation of the honeycomb lattices.
Starting from a uniform distribution of iABPs, we determine the average size of $B$ clusters and 
their growth for the various cut-off radii $R_v$. A particle is considered part of a cluster if the distance 
with another particle in the cluster is $r \leq 1.5 \sigma$. 
Figure \ref{fig:binary_cluster}(b) illustrates the average cluster growth. At short times $D_Rt < 0.1$, all 
systems exhibit a similar behavior independent of $R_v$. However, for longer times $D_R t > 1$, clusters 
are formed, which grow with time until they reach a stationary state. For the cutoff range $R_v=4\sigma$, 
the average cluster contains about four particles and there are many clusters 
(Fig.~\ref{fig:snapshot_honecomb}(a)). 
The clusters are very dynamic and reform continuously. The cluster size gradually increases with increasing 
vision cutoff range, reaching approximately $30$ iAPBs for $R_v=14\sigma$. It is important to note
that the coarsening effectively stops because the $A$ particles in the hexagon boundaries present an efficient 
barrier for the crossing of $B$ particles. Thus, the hexagonal structure is very stable despite its
dynamic nature.

\section{Summary and Conclusions}

We have studied binary mixtures of self-steering active particles, where particle motion and steering is based 
on the instantaneous position of neighboring particles using a minimal cognitive model. The particle speed, 
vision angle,  vision range, and sign and strength of maneuverability -- relative to their peers and the foreign 
species -- all play crucial roles in the self-organization and structure formation. 
Beyond a general overview of the enormous variability of emerging structures, we focus on the 
dynamical properties of three interesting self-organized phases, multimeric aggregates in 
off-stochiometric mixtures, 
prey-predator-type behavior, and the formation of honeycomb-type lattices. 

In the case of steering properties reminiscent of systems with electrostatic interactions,
the number ratio of the two components plays a crucial role in structure formation, along with vision angle 
and activity, which leads to the emergence of dimers, tetramers, and higher multimeric aggregates. 
For particle numbers which deviate slightly from stochiometric ratios, unpaired particle display a 
hopping dynamics between multimers, where they can knock out a particle in the existing aggregate and
substitute it, resulting in a sub-diffusive (caged) behavior at short times 
and diffusing motion at a longer time. The analysis of the mean hopping displacement shows a strong density 
dependence, where a high/low-density system has the smallest/highest hopping displacement.

In the case of a nonreciprocal steering response between the two particle types, 
i.e. $\Omega_{ab}=-\Omega_{ba}$,
and only one of the types (the ``prey") preferring the vicinity of their peers, the emergent dynamics displays
a predator-prey-like behavior. Our analysis reveals that the optimal angle for the predator to steer toward the 
prey is around $\pi/4$, corresponding to focused vision, which allows the predator to steer effectively 
toward the prey without getting distracted to much by other prey.
Our simulation results are strikingly similar to the behavior of some natural systems, like reef shark in 
a fish swarm \cite{carr_2015_shark}; this type of predator-prey behavior
has actually already been employed \cite{touahmi_2012_shark} to construct congestion avoidance
for multiple micro-robots. Our results should provide the necessary guidance to optimize the
design of such micro-robotic systems.

Finally, we considered a similar case with nonreciprocal $A$-$B$ interactions as before, but now its only the 
predators which look for the vicinity of their peers. This seemingly small change generated a completely 
different kind of self-organization, where predator clusters to form a very stable honeycomb lattice with the 
vision cuttoff range determining the size of the cluster and the lattice constant -- with larger clusters 
for larger vision range.

Our system displays some fundamental and phenomenological similarities with mixtures of chemically interacting 
particles, which produce or consume a chemical to which they are attracted or repelled \cite{Canalejo_2019_PRL}.
In both cases, nonequilibrium nonreciprocal interactions between particles, which break action-reaction symmetry
lead to new classes of active phase separation phenomena. For example, the formation of molecule-like binary
aggregates, and of honeycomb-like lattices are observed in both systems. However, a closer look reveals also 
several important differences. First, interactions in the chemical system are isotropic, while our iABPs have
directed sensing through the vision cone with a limited vision angle. Secondly, motion in the chemical systems
develops as a results of chemotactic motion, while ABPs move with constant speed; the latter implies that
iAPBs can only react to gathered information by steering, for which we impose a limited maneuverability. 
These differences imply strikingly different
emergent structure formation and dynamics. For example, hopping ``defects'' in the phase of small
molecule-like aggregates in only seen for iABPs, as well as the animal-like predator-prey behavior. In 
contrast, static cluster of one particle type propelled by corona and tail of the other particle type are
only seen in the chemical system.

A important advantage of our "minimalist" model of binary cognitive particle mixtures is its 
flexibility, which facilitates the description of a large variety of natural \cite{tsang_2020_AIS} 
and artificial \cite{touahmi_2012_shark, chen_PRL_2024} systems.
Here, (micro)robotic systems are very promising experimental model systems, as they allow the 
implementation of many different, simple or complex, interaction and steering rules. For example,
a recent study \cite{chen_PRL_2024} of a binary systems of programmable robots with non-reciprocal interactions, 
where species $A$ aligns with $B$ but $B$ antialigns with $A$ demonstrates the emergence of a collective 
chiral motion that can be stabilized by limiting the robot angular speed to be below a threshold. 
It will also be interesting to explore and utilize possible synergies between agent-based models
and Cahn-Hillard-type continuum field theoretical approaches 
\cite{fruchart_2021_Nature, suchanek_PRL_2023, kreienkamp_PRL_2024}
to elucidate the behavior of non-reciprocal multi-component systems.

Thus, the extension of single-component cognitive active particle 
systems to nonequilibrium two-component mixtures with nonreciprocal interactions generates an
enormous richness of emergent complexity and variability of self-organization and dynamical 
scenarios. Despite of several recent studies, the large space of of self-organization 
behavior has just been scratched at the surface so far, and more detailed studies to elucidate 
the underlying physical mechanisms are required.

\section*{Data Availability}
The data that support the findings of this paper are openly
available \cite{DATA}.


%

\end{document}